\documentclass[journal]{vgtc}                % final (journal style)
\ifpdf%                                % if we use pdflatex
  \pdfoutput=1\relax                   % create PDFs from pdfLaTeX
  \usepackage{graphicx}                % allow us to embed graphics files
  \DeclareGraphicsExtensions{.pdf,.png,.jpg,.jpeg} % for pdflatex we expect .pdf, .png, or .jpg files
\else%                                 % else we use pure latex
  \usepackage{graphicx}                % allow us to embed graphics files
  \DeclareGraphicsExtensions{.eps}     % for pure latex we expect eps files
\fi%

\graphicspath{{figures/}{pictures/}{images/}{./}} % where to search for the images

\usepackage{microtype}                 % use micro-typography (slightly more compact, better to read)
\PassOptionsToPackage{warn}{textcomp}  % to address font issues with \textrightarrow
\usepackage{textcomp}                  % use better special symbols
\usepackage{mathptmx}                  % use matching math font
\usepackage{times}                     % we use Times as the main font
\usepackage{cite}                      % needed to automatically sort the references
\usepackage{tabu}                      % only used for the table example
\usepackage{booktabs}                  % only used for the table example
\usepackage{caption}
\onlineid{0}

\vgtccategory{system}
\vgtcpapertype{poster}

\title{regvis.net -- A Visual Bibliography of Regulatory Visualization}

\author{Zhibin Niu*, Runlin Li, Junqi Wu, Yaqi Xue, Jiawan Zhang}
\authorfooter{
Zhibin Niu, Runlin Li, Junqi Wu, Yaqi Xue, and Jiawan Zhang are with College of Intelligence and Computing, Tianjin University. Corresponding: zniu@tju.edu.cn
}

\shortauthortitle{Zhibin Niu \MakeLowercase{\textit{et al.}}: regvis.net -- A Visual Bibliography of Regulatory Visualization}
\abstract{Information visualization and visual analytics technology has attracted significant attention from the financial regulation community. In this research, we present regvis.net, a visual survey of regulatory visualization that allows researchers from both the computing and financial communities to review their literature of interest. We have collected and manually tagged more than 80 regulation-visualization related publications. To the best of our knowledge, this is the first publication set tailored for regulatory visualization. We have provided a webpage (http://regvis.net) for interactive searches and filtering. Each publication is represented by a thumbnail of the representative system interface or key visualization chart, and users can conduct multi-condition screening explorations and fixed text searches.
} % end of abstract


\CCScatlist{
  \CCScatTwelve{Human-centered computing}{Visualization}{Visualization application domains}{Visual analytics};
}

\teaser{
  \includegraphics[width=1\linewidth]{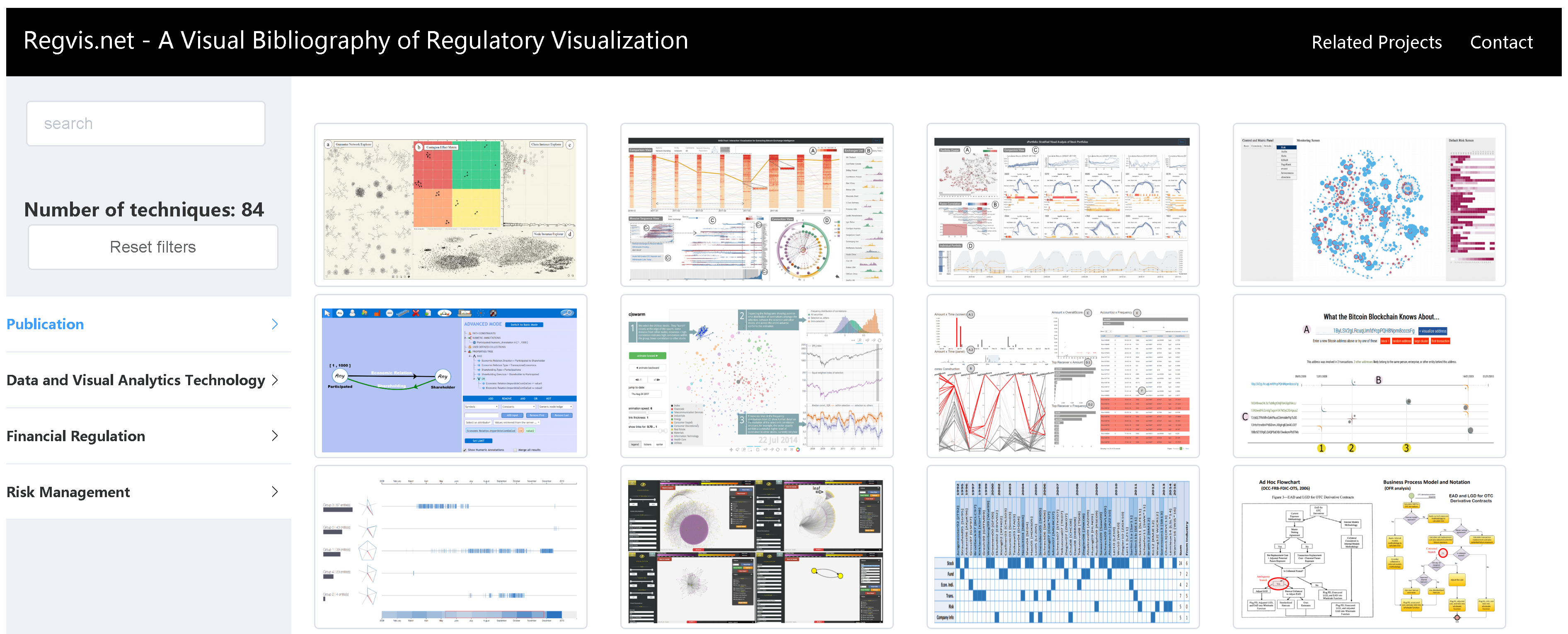}
  \caption{Visualization interface for http://www.regvis.net. Users can choose to filter for their publications of interest through drop-down menus (publication, data and visual analytics technology, financial regulation, and risk management) in the left panel. Each publication is represented by a thumbnail of the main system interface view or key visualization view.}
  \label{interface}
}

\vgtcinsertpkg

\begin{document}


\firstsection{Introduction}

\maketitle
Regulatory technology (RegTech) is an emerging subfield of financial technology (FinTech) designed to improve transparency and consistency and address regulatory challenges faced by financial services providers, including monitoring, reporting, and compliance obligations. Regulatory technology is reshaping risk management and changing the nature of financial markets and services through data analytics, predictive modeling, and statistical tools~\cite{arner2016fintech, arner2017fintech}. Although regulatory technology was initially seen as a compliance solution, it also has potential as a risk mitigation tool, mainly because of the relationship between compliance and risk management functions. The rapid development of regulatory technology has raised awareness of information visualization, visual analytics, and artificial intelligence in this area. Andy Haldane, the chief economist and the Executive Director of Monetary Analysis and Statistics at the Bank of England, imaging the future of data visual analysis in the field of regulatory technology during a keynote address at Birmingham University as~\cite{haldane2014managing}:

\begin{quote}
  \emph{I have a dream. It is futuristic, but realistic. It involves a Star Trek chair and a bank of monitors. It would involve tracking the global flow of funds in close to real time (from a Star Trek chair using a bank of monitors), in much the same way as happens with global weather systems and global internet traffic. Its centerpiece would be a global map of financial flows, charting spill-overs and correlations.}
\end{quote}

We define regulatory visualization (RegVis) as information visualization and visual analytics for regulatory technology. We present http://www.regvis.net, a web-based visual bibliography of regulatory visualization that serves as a comprehensive collection of published work on visualization and visual analytics in regulatory technology. Through 1 July 2020, we manually collected over 80 publications on the topic of regulatory visualization technology. The sources of these publications are diverse. Some are from economics management and financial journals such as Journal of Financial Stability, Journal of Banking and Finance, or European Central Bank Working Papers, while others are from information visualization-related conferences or journals such as IEEE Conference on Visual Analytics Science and Technology and IEEE Transactions on Visualization and Computer Graphics. We have provided this platform to attract attention from both the financial and computing communities, as we believe that interdisciplinary collaboration will foster new research opportunities.

\begin{figure*}[htp]
    \centering
    \captionsetup{justification=centering,margin=2cm}
    \includegraphics[width=\textwidth]{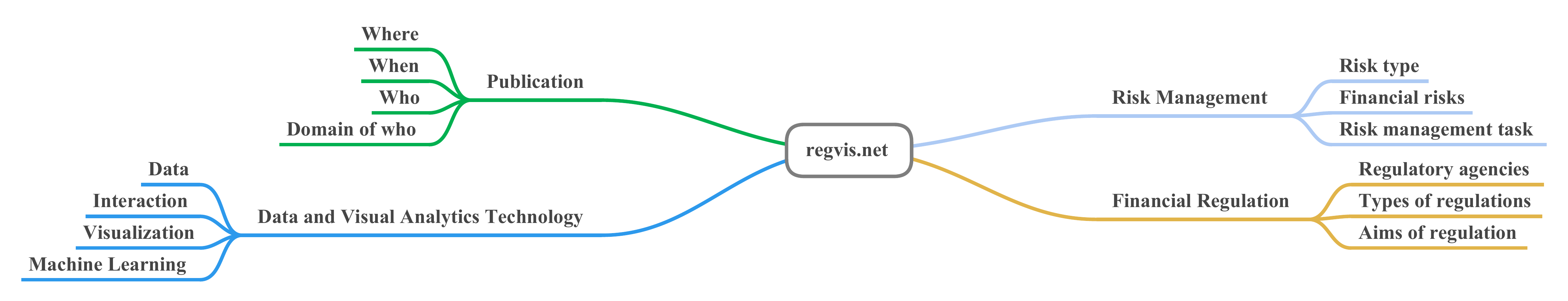}
    \caption{Top three levels of the taxonomy for regulatory visualization.} \vspace{-10pt}
    \label{taxonomy}
\end{figure*}

The regvis.net visual survey is inspired by previous work such as treevis~\cite{schulz2011treevis}, textvis~\cite{kucher2014text}, financevis~\cite{dumas2014financevis}, and others. As with these earlier efforts, regvis.net is essentially a filter-supported website that enables users to screen for various factors. Our regvis.net is in the same application domain as financevis.net. The difference is that our focus is more on visualization related works in the regulatory technology field. We did not include visualization related publications in other financial technology fields such as intelligent investment and intelligent operations. Another significant difference is that many of these previous efforts used the novel visualization chart as a representative image, while we used the system’s visual interface. The reason is that often such tools are used to resolve sophisticated analytical tasks, with the goal most often achieved through coordinated interactions between functional views. For such complicated interdisciplinary research, functional and coordinated system-level visualization and interactive design are essential.

\section{Taxonomy of Regvis.net}

As mentioned above, RegTech is an extended form of financial risk management. We worked with financial experts to organize a taxonomy of regulatory visualization that is the basis of our visual survey. We have included only the top three levels of the taxonomy in Figure~\ref{taxonomy}, due to limited space. Readers should refer to http://regvis.net for more information. In detail, the taxonomy includes the following.

\textbf{Publication} The source of a publication is usually the primary factor in judging a work’s quality, impact, and level of interest for researchers. The bibliography filters by conference or journal, date of publication, author(s), and other aspects.

\textbf{Data and Visual Analytics Technology} RegTech-related research usually spans a broad field of technology, including data mining, machine learning, information visualization, and visual analytics. We designed and provided a series of fine-grained filters reflecting these forms of technology. There are four subcategories: data, machine learning methods, visualization methods, and interactions. The data category specifies the type of dataset and data source (such as open-source data or dataset from banks). Machine learning technology and visualization methods are summarized from publications. The taxonomy of interactions follows the seminal publications of~\cite{shneiderman1996eyes,leung1994review, yi2007toward,dumas2014financevis}.

\textbf{Financial Regulation} The financial regulation and risk management categories are designed to address RegTech problems. The taxonomy of financial regulation, according to~\cite{finreg}, includes three subcategories: regulatory agencies, types of regulation, and regulatory goals.

\textbf{Risk Management} Financial regulation and risk management are closely related. One of the main differences is that financial regulation deal more with financial regulatory services, while risk management focuses on RegTech and risk management-related frameworks, methodologies, and standards. This category includes three subcategories~\cite{iso2009risk,purdy2010iso,mcneil2015quantitative}: risk type, financial risk, and risk management tasks.

\section{Visual Interface of Regvis.net}

The visual interface of regvis.net is composed of three parts (see Figure~\ref{interface}). (1) The main view is a collection of thumbnails of publications. The works are ordered by their publication date to facilitate readers tracing research trends. (2) The panel view on the left is a series of interactive filters for the taxonomy described in the previous section. It supports multi-condition screening exploration through a series of drop-down menus and a topic search function. (3) Related projects and contact buttons on the navigation panel list other visual survey projects and related publications.

\section{Conclusion}
In this research, we introduce regulatory visualization and present http://regvis.net, a visual bibliography for interdisciplinary research. The main contribution of this work is the RegVis related taxonomy. Based on that framework, we collected and manually tagged over 80 related publications. To our knowledge, this is the first dataset of publications gathered on this topic. We hope the collection and interactive web-based application will help foster potential research opportunities in both the computing and financial fields. It should be noted that the RegTech and RegVis taxonomy is regularly evolving, as these are emerging topics. We will continue to update the taxonomy and adjust publication tags. We will also continue to include new publications in this web tool so that experts will be better able to keep abreast of new developments in the field of regulatory visualization. A comprehensive review of this topic is also in production.

\acknowledgments{
This work was supported by NSFC (61802278) and the MOE Key Laboratory Foundation for AI (AI2019004).}

\bibliographystyle{abbrv-doi}

\end{document}